\documentclass[12pt]{iopart}

\usepackage{iopams}
\usepackage{latexsym}
\usepackage{graphicx}

\newcommand{\text}[1]{{\rm{#1}}}

\def\bh#1{black hole#1 (BH#1)\gdef\bh{BH}} 
\def\ahz#1{apparent horizon#1 (AH#1)\gdef\ahz{AH}} 
\def\bbh#1{binary black hole#1 (BBH#1)\gdef\bbh{BBH}} 
\def\qnm#1{quasi-normal mode#1 (QNM#1)\gdef\qnm{QNM}}
\def\isco#1{innermost stable circular orbit#1 (ISCO#1)\gdef\isco{ISCO}} 
\def\lisa#1{Laser Interferometer Space Antenna#1 (LISA#1)\gdef\lisa{LISA}} 
\def\kickrange{25-82\,km/s}
\def\madm{M_\text{ADM}}
\newcommand\iu{i}  

\begin{document}


\title{Unequal Mass Binary Black Hole Plunges and Gravitational Recoil}


\author{Frank~Herrmann}
\address{Center for Gravitational Wave Physics\\
Institute for Gravitational Physics and Geometry\\
Penn State University, University Park, PA 16802}

\author{Ian~Hinder}
\address{Center for Gravitational Wave Physics\\
Institute for Gravitational Physics and Geometry\\
Penn State University, University Park, PA 16802}

\author{Deirdre~Shoemaker}
\address{Center for Gravitational Wave Physics\\
Institute for Gravitational Physics and Geometry\\
Penn State University, University Park, PA 16802}

\author{Pablo~Laguna}
\address{Center for Gravitational Wave Physics\\
Institute for Gravitational Physics and Geometry\\
Penn State University, University Park, PA 16802}


\date{Dec 30, 2006}


\begin{abstract}
  We present results from fully nonlinear simulations of unequal mass
  binary black holes plunging from close separations well inside the
  innermost stable circular orbit with mass ratios $q \equiv M_1/M_2 =
  \lbrace 1,0.85,0.78,0.55,0.32 \rbrace$, or equivalently, with
  reduced mass parameters $\eta \equiv M_1M_2/(M_1+M_2)^2 = \lbrace
  0.25, 0.248, 0.246, 0.229, 0.183 \rbrace$. For each case, the
  initial binary orbital parameters are chosen from the Cook-Baumgarte
  equal-mass ISCO configuration. We show waveforms of the dominant
  $\ell=2,3$ modes and compute estimates of energy and angular
  momentum radiated. For the plunges from the close separations
  considered, we measure kick velocities from gravitational radiation
  recoil in the range \kickrange. Due to the initial close separations
  our kick velocity estimates should be understood as a lower
  bound. The close configurations considered are also likely to
  contain significant eccentricities influencing the recoil velocity.
\end{abstract}




\maketitle


\section{Introduction}

The \lisa{} will offer a unique look at merging supermassive \bh{s}
~\cite{2003AdSpR..32.1233D,2003AAS...202.3701P}.
When galaxies collide, the \bh{s} at the center of
each galaxy~\cite{1998Natur.395A..14R,Magorrian:1997hw}  
will merge and radiate gravitational waves. In a generic
merger situation, the colliding \bh{s} will have different masses and
spins, thus they will radiate gravitational waves anisotropically. This
anisotropic radiation will carry both a net angular and 
linear momentum~\cite{1973ApJ...183..657B}.
The linear momentum emitted  
implies a recoil of,  or kick to, the \bh{} product of the merger.
A large enough recoil may cause the \bh{} resulting from the merger
to be strongly offset from the
center of the galaxy or potentially even kicked out of small 
galaxies~\cite{2004ApJ...606L..17M,Merritt:2004xa}.
Such a scenario would have a significant impact on the standard picture 
of merger tree history of galaxies.

A complete investigation of the gravitational recoil from unequal mass \bh{} mergers 
must include the contributions to the kick from the inspiral, merger and ringdown
stages of the life of the binary. 
Given the tremendous progress that numerical relativity 
has recently made in simulating \bbh{}
systems~\cite{Bruegmann:2003aw,Alcubierre2003:pre-ISCO-coalescence-times,
Pretorius:2005gq,Campanelli:2005dd,Baker:2005vv,Diener:2005mg}, 
fully numerical investigations can now be used to investigate gravitational recoil. Recent numerical studies~\cite{Baker:2006vn,Gonzalez:2006md} as well as 
post-Newtonian 
estimates\cite{Blanchet:2005rj,Damour:2006tr} suggest that 
most of the kick gets accumulated during the plunge. 
Motivated by these observations, 
we carried out a series of unequal mass \bbh{s} simulations,
using initial data that corresponds to \bh{s} 
plunging from \isco{}, i.e. initial separations of $~ 2.34M$.
In particular,
we focus on the dominant $\ell=2,3$ modes and study
how the energy radiated in these modes changes as the mass ratio is
varied.  Additionally, we calculate the energy and angular momentum
radiated, and from the gravitational radiation we estimate kick
velocities in the range \kickrange\ acquired by the final \bh{.}

Soon after the completion of our study, Baker et
al.~\cite{Baker:2006vn} studied a much further separated system and
Gonzales et al.~\cite{Gonzalez:2006md} published fully relativistic
kick estimates from \bh{s} with initial separation of $7M$ for a broad
range of mass ratios, from $q=M_1/M_2=1$ to $q =0.25$, or equivalently
$\eta = q/(1 + q)^2$ from 0.25 to 0.16.  They estimated a maximum kick
of $175.70$ km\,s$^{-1}$ for a mass ratio of $\eta =
0.195$. Using close-limit approximation techniques Sopuerta et
al. investigated gravitational wave recoil and also the effect of
small eccentricities~\cite{Sopuerta:2006wj,Sopuerta:2006et}.


\emph{Initial Data.} The initial data sets are constructed via the
puncture method using a spectral code~\cite{Brandt97b,Ansorg:2004ds}.
The essence of this method is to solve the Hamiltonian constraint for
the conformal factor $\phi$. The initial three-metric is conformally
flat, maximally sliced, and the extrinsic curvature is given by the
Bowen-York solution to the momentum constraint. The conformal factor
$\phi$ is used to set the initial lapse as
$\alpha=\phi^{-2}$~\cite{Alcubierre02a,Campanelli:2005dd}, while the
initial shift is $\beta^i=0$.

For the equal mass setup, we evolve the so-called QC-0 initial data
set~\cite{Baker:2002qf}.  This is intended to represent a
quasi-circular configuration of inspiraling puncture \bbh{s} at the
\isco{}. QC-0 data have been used as the starting point by other
studies~\cite{Alcubierre2003:pre-ISCO-coalescence-times,Campanelli:2005dd,Baker:2005vv}.
The \bh{s} in QC-0 perform about half of an orbit prior to
merging~\cite{Campanelli:2005dd,Baker:2005vv}; that is, QC-0 looks
more like a plunge/grazing collision. The intersection of the event
horizon with the initial slice, however, has the topology of two
separate spheres~\cite{Alcubierre2003:pre-ISCO-coalescence-times}.

The puncture \bbh{} data of the Bowen-York type are defined by
the bare masses $m_{1,2}$ of the \bh{s,} their coordinate locations
$C_{1,2}$, assumed to be along the $x$-axis in the $xy$-plane, and
their linear momentum $P_{1,2}$, pointing along the $y$-axis.  In the
construction of the initial data, we vary $m_2 =
\lbrace{1,1.2,1.3,2,4\rbrace}\,m_1$ while keeping $m_1$ fixed to its
QC-0 value of $0.453$.  We also keep fixed to the QC-0 values the
puncture coordinate separation $d \equiv |C_1-C_2|=2.34$ and the
momentum parameters $P_{1,2} = \pm 0.333$, which means that the
angular momentum value, $J= 0.779$, also remains unchanged. Note that
the total ADM mass, $\madm$, of the configurations does change as do
$J$ and $P$ when given in ADM mass units. Due to this setup, our
initial data of unequal mass \bbh{} systems do not obey the
quasi-circular orbit condition of minimal binding
energy~\cite{1994PhRvD..50.5025C}.  The present work is aimed at
investigating the effects of varying, in the QC-0 set-up, the bare
mass parameter only.  The motivation behind this choice was to study
gravitational recoil starting from the \isco{} plunge with the
simplest parameter exploration. There is strong indication that the
kick imparted to the \bh{} that has resulted from the merger is
dominated by the gravitational recoil during the plunge from \isco{}
\cite{Blanchet:2005rj}.  We are currently extending the present study
and investigating both unequal mass \bbh{} mergers with initial
separations outside of \isco{} in quasi-circular orbit and plunge
configurations with Post-Newtonian orbital parameters
\cite{Blanchet:2005rj}.

Table~\ref{tbl:IDtable} summarizes the parameters in our simulations.
We list the total mass $M=M_1+M_2$, the mass ratio parameter $\eta=M_1
M_2/M^2$ where $M_{1,2}$ are the irreducible masses of the \bh{s}
computed from their individual \ahz{} areas, as well as angular
momentum $J$ and momentum parameter $P$ in terms of the reduced mass,
$\mu=M_1M_2/M$. Table~\ref{tbl:IDtable} also provides the time
$t_\text{AH}$ in $\madm$ units when a common \ahz{} is first found.
For QC-0 the orbital period of the equal mass case is estimated as
$t=37.4\,\madm$.  The drop in the time to merger $t_\text{AH}$ in our
results should not be interpreted as ``unequal mass binaries merge
faster''. The effect is due to our approach in which the angular
momentum of the configuration decreases as the bare mass ratio is
decreased.  Ideally one would like to compare initial data sets which
are far separated and have the same orbital frequency.  It is possible
that for our cases with $q\le0.55$, a common event horizon already
exists in the initial data slice; and, therefore, we are evolving a
single, distorted \bh{.}

\begin{table}
  \begin{center}
  \begin{tabular}{cccccccc}
  $q$  &  $M_1/M$ & $M_2/M$ & $M$ & $\eta$ & $P/\mu$ & $J/(\mu M)$ & $t_\text{AH}$ \\
  \hline 
  1.00& 0.512& 0.512& 1.024& 0.250& 1.290& 2.948& 18.4\\ 
  0.85& 0.482& 0.565& 1.047& 0.248& 1.180& 2.637& 12.2\\ 
  0.78& 0.463& 0.597& 1.060& 0.246& 1.122& 2.476& \phantom{0}9.9\\ 
  0.55& 0.391& 0.712& 1.103& 0.229& 0.933& 1.979& \phantom{0}5.5\\ 
  0.32& 0.277& 0.872& 1.149& 0.183& 0.692& 1.409& \phantom{0}2.5\\ 
  \end{tabular}
  \end{center}
  \caption{Initial data parameters and properties: The $q=1$ case is
    the QC-0 data set in~\cite{Baker:2002qf}. All models have puncture
    bare mass $m_1=0.453$, momenta $P_{1,2}=\pm 0.333$ and coordinate
    separation $d=2.34$.  The total mass $M$ is given in $\madm$ units 
    as well as the coordinate time $t_\text{AH}$ at which a common apparent
    horizon was found.}
\label{tbl:IDtable}
\end{table}


\emph{Methods.}  The evolutions were carried out using a code that
solves the BSSN 3+1 formulation of Einstein's
equations~\cite{Nakamura87, Shibata95, Baumgarte99}. We use the
``moving punctures'' approach without
excision~\cite{Baker:2005vv,Campanelli:2005dd}. The code was produced
by the \texttt{Kranc} code generation package~\cite{Husa:2004ip} and
uses the \texttt{Cactus} infrastructure. The simulations were
performed using fourth order accurate centered finite differencing,
except for the advection terms which were one-sided and second order
accurate.  The temporal updating is carried out with a three-step
iterative Crank-Nicholson scheme with Courant factor of 0.25.  Tests
in the waveform for the equal-mass setup using resolutions $h=\lbrace
1/16, 1/20, 1/32\rbrace$ produced convergence slightly below second
order.  Mesh refinement in the code is provided by
\texttt{Carpet}~\cite{Schnetter-etal-03b}, and tracking of \ahz{s} is
handled by {\tt AHFinderDirect}~\cite{Thornburg2003:AH-finding}.

The gauge conditions used were modified versions of the 1+log lapse
and $\Gamma$-driver shifts.  Specifically, the lapse $\alpha$ was
evolved using $\partial_t \alpha = - 2\alpha K$, where $K$ is the
trace of the extrinsic curvature.  On the other hand, the shift vector
was obtained from~\cite{Baker:2005vv}: $\partial_t \beta^i = F B^i$
and $\partial_t B^i = \partial_t \tilde{\Gamma}^i
-\beta^j\partial_j\tilde{\Gamma}^i- \xi B^i$ with $\xi$ a constant
dissipative parameter and $F=3 \alpha/4$, which guarantees that the
asymptotic gauge speed associated with the longitudinal shift
components is equal to the speed of light. The evolutions were started
with $\beta^i$ and $B^i=0$. The advection term
$\beta^j\partial_j\tilde{\Gamma}^i$ removes certain zero-speed modes
of the system as analyzed in~\cite{vanMeter:2006vi}. The parameter
$\xi$ can be used to tune the rate of horizon expansion over the
course of the evolution; large values lead to faster horizon
growth. We have used values in the range $2\le \xi \le 5$ and have not
found any instabilities in this range. For the runs reported in this
paper we used $\xi = 4$.

Our computational domain consisted of fixed 2:1 mesh refinements with
5 levels.  The finest grid spanned $-2 \le x,y \le 2$ and $0 \le z \le
2$, with the coarsest $-96 \le x,y \le 96$ and $0 \le z \le 96$ (we
use bitant symmetry in the $z$ direction). All the refinement levels
except for the coarsest have the same number of grid points. We have
run the unequal \bbh{} models at two different resolutions, $h=
1/16,1/20$.


\begin{figure}
\begin{center}
\includegraphics{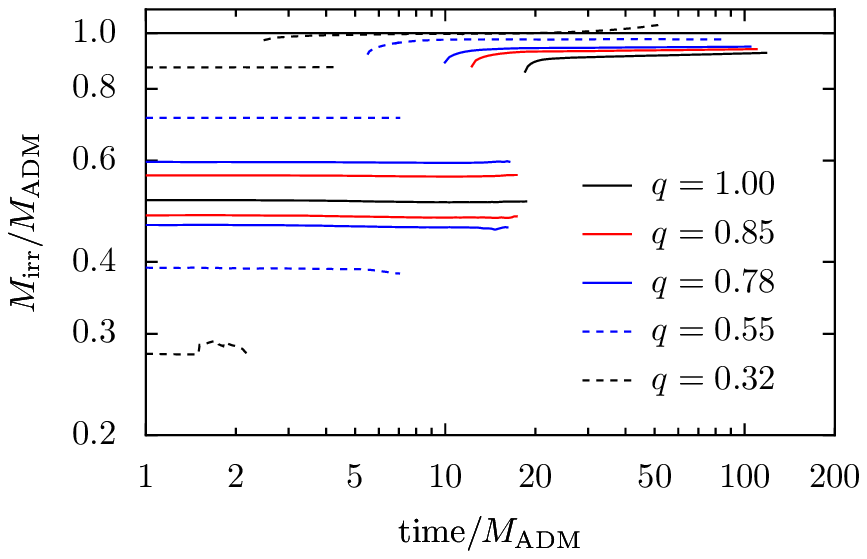}
\end{center}
\caption{The irreducible mass of the apparent horizon as a function of
  time for different mass ratios $q$.}
\label{fig:horizon_masses}
\end{figure}

\emph{Results.}  Fig.~\ref{fig:horizon_masses} shows the irreducible
mass of the \ahz{s} as a function of time, where
$M_{irr}=\sqrt{A/16\pi}$ and $A$ is the \ahz{} area.  In all
simulations with $q>0.32$, we are able to track very accurately
$M_{irr}$ for both of the individual merging \bh{s}, as well as the
resulting \bh{}, over the entire course of the simulation.  In the
case of $q=0.32$, we cannot track the smaller individual irreducible
mass very accurately before merger.  There is also a spurious drift in
the common apparent horizon mass beyond the initial ADM mass content
of the spacetime but the error only grows about 2\% over
$t=50\,\madm$.  Notice that the smaller the value of $q$, the earlier
the merger as indicated by $t_{AH}$ in Table~\ref{tbl:IDtable}.
Post-merger, the individual \ahz{s} loose their meaning as apparent
horizons since by definition the \ahz{} denotes the \emph{outermost}
marginally trapped surfaces. The individual surfaces remain marginally
trapped.

\begin{figure}
\begin{center}
\includegraphics{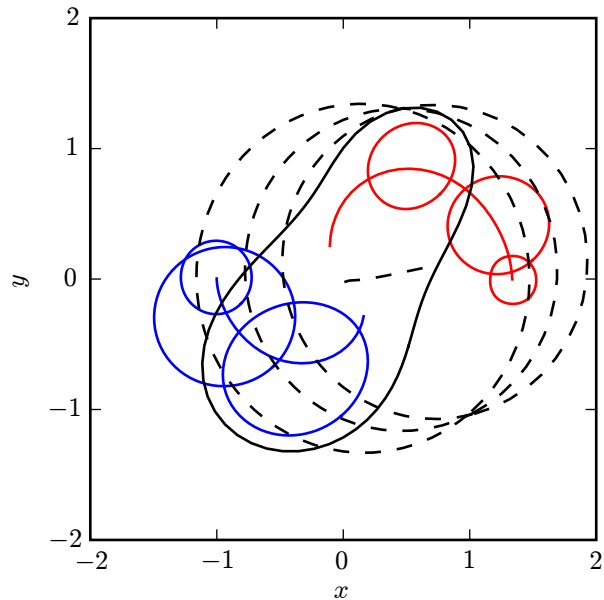}
\end{center}
\caption{Snapshots of the \ahz{} location in the $xy$-plane for the
  case $q=0.78$. The larger \bh{} is on the left moving toward the
  bottom. The snapshots are taken every $4.6\,\madm$ prior to merger
  and at $t=\lbrace 40,80,105 \rbrace\,\madm$ after merger. The first
  common \ahz{} is found at $t=9.9\,\madm$. The trajectories of the
  \ahz{} centroids are also shown. The common \ahz{} moves to the
  right and slightly upward after merger.}
\label{fig:horizon_snapshots}
\end{figure}

Fig.~\ref{fig:horizon_snapshots} shows snapshots of the horizons every
$4.6\,\madm$ before merger and at $t=\lbrace 40,80,105 \rbrace\,\madm$
after merger for the $q=0.78$ case. The other cases are qualitatively
similar. Notice how the initial common \ahz{} has an asymmetric peanut
shape due to the unequal masses.  Soon after it appears, the common
\ahz{} becomes spherical, as the dynamical gauges drive the
coordinates toward those of a single \bh{.} At that moment, the common
\ahz{} begins to drift slowly away from the origin. The last \ahz{}
snapshot in Fig.~\ref{fig:horizon_snapshots} was taken at
$t=105\,\madm$. The evident drift in the coordinate location of the
common \ahz{} provides a \emph{hint} that a kick is generated as a
consequence of gravitational recoil.

\begin{figure}
\begin{center}
\includegraphics{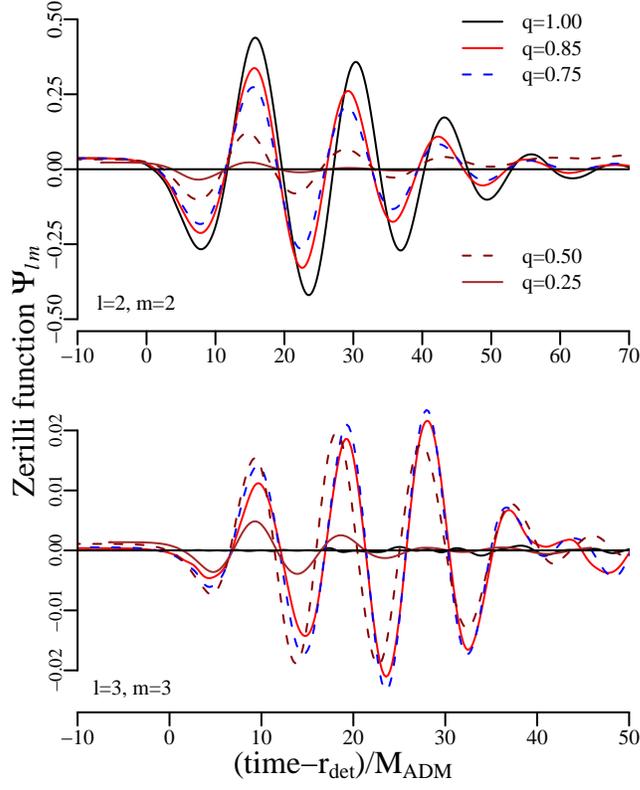}
\end{center}
\caption{The dominant modes ($\ell=2, m=2$ and $\ell=3, m=3$) of the
  real part of the Zerilli function $\psi_{\ell m}$ against time for
  the different $q$ ratios. The waveforms were extracted at $r=15$.
  The $\ell=2, m=2$ mode decreases in amplitude with decreasing $q$
  while the $\ell=3,m=3$ mode increases and then decreases again.}
\label{fig:waveforms}
\end{figure}

For waveform extraction, we compute Zerilli modes $\psi_{\ell m}$
using the Abrahams \& Price convention~\cite{2005CQGra..22R.167N}.
Formally, the method assumes a spherically symmetric background.
Simulations of QC-0 data~\cite{Campanelli:2005dd,Baker:2005vv} have
produced rotating \bh{s} with Kerr parameter $a\sim0.7$. The main
effect of using Zerilli extraction in spacetimes with significant
angular momentum content will be a spurious amplitude offset in the
waveform~\cite{2001PhRvL..87A1103A,Brandt:1994db}.
Fig.~\ref{fig:waveforms} displays the dominant modes ($\ell=2,m=2$ and
$\ell=3,m=3$) for the different $q$ simulations. The extraction
surface is located at $r=15$ with the outer boundary at $r=96$.
Notice in Fig.~\ref{fig:waveforms} that the $\ell=2, m=2$ mode simply
decreases in amplitude as the $q$ ratio is decreased. This is mostly
due to the initial data approaching that of a single distorted \bh{.}
The $\ell=3, m=3$ mode is smallest for the equal mass case (where it
should be 0) and for the $q=0.32$ case.  Around $t=50\,\madm$, the
lower amplitude waveforms become affected by outer boundary
effects. The strongly dominant $\ell=2, m=2$ mode remains accurate
until the end of the simulation except for the $q=0.32$ case where
even the dominant mode is affected by the boundary by $t=60\madm$.

\begin{table}
  \begin{center}
  \begin{tabular}{cccccc}
  $q$   &  $E_{\text{rad}}$      & $E_0$     & $E_2$     & $J_{\text{rad}}$     & $V\,(\text{km/s})$ \\
  \hline 
  1.00 &    $2.7\pm 0.8$  & $0.05\pm 0.1$    & $1.3\pm 0.4$     &  $15\pm 6$     &  $1\pm 2$   \\
  0.85 &    $1.7\pm 0.2$  & $0.038\pm 0.004$  & $0.78\pm 0.02$   &  $10\pm 0.8$  &  $50\pm 20$ \\
  0.78 &    $1.1\pm 0.8$  & $0.06\pm 0.04$    & $0.55\pm 0.02$   &  $7.4\pm 0.8$ &  $72\pm 40$ \\
  0.55 &    $0.4\pm 0.2$  & $0.011\pm 0.006$ & $0.16\pm 0.06$   &  $2.6\pm 0.6$  &  $90\pm 50$ \\
  0.32 &    $0.05$        & $0.001$           & $0.024$          &  $0.4$        &  $31$  \\
  \end{tabular}
  \end{center}
  \caption{Estimates of radiated energy $E_{\text{rad}}$ (total), $E_0$ ($m=0$
    mode) and $E_2$ ($m=2$ mode) as a \% of the total initial ADM
    energy.  Radiated angular momentum $J_{\text{rad}}$ as a \% of the
    initial angular momentum.  Kick velocities $V$ from gravitational
    recoil. The errors are the differences between the $h=1/16$ and
    1/20 resolution results multiplied by 2.}
  \label{tbl:Kickstable}
\end{table}

From the extracted modes, Table~\ref{tbl:Kickstable} summarizes
estimates of the energy $E_{\text{rad}}$ and angular momentum
$J_{\text{rad}}$ emitted as \% of the total energy and angular
momentum as well as the recoil velocities $V$ in
km/s~\cite{Wiseman:1992dv}. We compute the radiated energy
from~\cite{2005CQGra..22R.167N}
\begin{equation}
\frac{dE_{\text{rad}}}{dt} = \frac{1}{32\pi} \sum_{\ell,m} \left[\left|\frac{d\psi^+_{\ell m}}{dt}\right|^2+\left|\psi^\times_{\ell m}\right|^2\right] \,,
\end{equation}
the radiated angular momentum via
\begin{equation}
\frac{dJ_{\text{rad}}}{dt} = \frac{1}{32\pi} \sum_{\ell,m} \iu m \left[\frac{d\psi^+_{\ell m}}{dt}\left(\psi^+_{\ell m}\right)^\ast + \psi^\times_{\ell m} \int_{-\infty}^t \left(\psi^\times_{\ell m}\right)^\ast dt^\prime\right]
\end{equation}
and the recoil velocity from 
\begin{eqnarray*}
h_+ -\iu h_\times &= \frac{1}{\sqrt{2}r}\sum_{\ell, m} \left[\psi_{\ell m}^+ -\iu \int_{-\infty}^t \psi_{\ell m}^\times dt^\prime\right]{}_{-2}Y^{\ell m}(\theta,\phi)\\
\frac{dP^k}{dt} &= \frac{r^2}{16\pi} \int \left[\left(\frac{dh_+}{dt}\right)^2+\left(\frac{dh_\times}{dt}\right)^2\right]n^k  d\Omega\,.
\end{eqnarray*}
Here ${}_{-2}Y^{\ell m}(\theta,\phi)$ denotes the spin-weight $-2$
tensor spherical harmonics. The kick velocity is obtained from
$M\,V^i=\int P^i\,dt$ with $M$ the total mass of the binary.  The
reported recoil velocity is $|V|$.

As expected, the radiated energy and angular momentum are strongly
dominated by the $\ell=2, m=2$ mode.  The energy and angular momentum
radiated for the $q=1$ case are in good agreement with previous work
~\cite{Baker:2005vv,Campanelli:2005dd}. The numbers reported in the
table are computed from the $h=1/20$ simulations. The errors are the
difference between the quantities computed from the $h=\lbrace 1/16,
1/20\rbrace$ resolution simulations multiplied by 2. The Richardson
error estimate for our resolutions and 2nd order convergence is 16/9
and 2 is used as a conservative bound.  We do not report errors for
the $q=0.32$ because the $h=1/16$ simulation has insufficient
resolution to compute the waveform. With increasing resolution the
estimates of $E_{\text{rad}}$ and $J_{\text{rad}}$ did increase, so
assuming convergence, one would expect the continuum estimates for
radiated energies and angular momenta to be slightly higher than those
reported here.

In order to compare the gravitational wave content of different
simulations, Table~\ref{tbl:Kickstable} also presents the energy
emitted in the $m=0$ and $m=2$ modes, i.e. $E_0$ and $E_2$
respectively, as a \% of the total initial ADM energy.  The
$\ell=2, m=0$ waveform in the $q=1$ case changes quite strongly between
the $h=1/20$ and $h=1/16$ simulations, resulting in large error bars
on $E_0$ and $V$ (which should be zero for the equal mass case). Since
the effect is not as strong for the $\ell=2, m=2$ mode, the dominant
mode for $E$ and $J_{rad}$, the error bars are smaller.  The
$\ell=2, m=2$ mode dominates for a pure inspiral, whereas the
$\ell=2, m=0$ mode dominates for a plunge, so we would expect a larger
$E_2/E_0$ ratio for inspirals than for plunges. It is reassuring that
differences in the radiated energies of these modes exist; however,
due to the construction of the initial data sequence, these
differences do not reflect a change of $q$ only. It will be
interesting to see how this changes with further separated, truly
inspiraling configurations starting from outside ISCO.

\begin{figure}
\begin{center}
\includegraphics{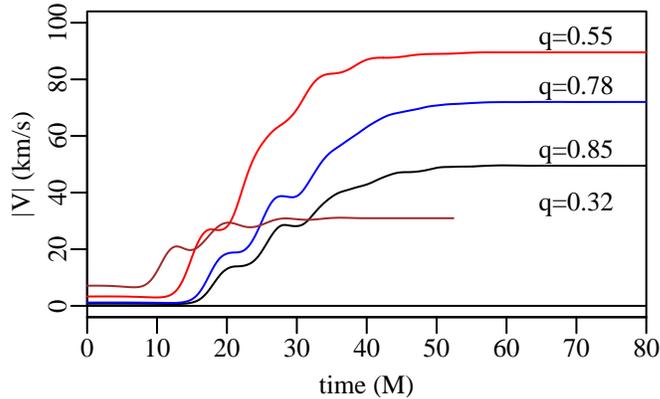}
\end{center}
\caption{Recoil velocity accumulated against time for the different
  models. Note that in comparison to far separated inspiral models
  (\cite{Baker:2006vn,Gonzalez:2006md}) the inspiral contribution is
  missing. Also the $q=0.32$ case shows an initial offset from zero
  which comes from the too close extraction radius.}
\label{fig:recoil_vs_time}
\end{figure}

The recoil velocities, $V$, reported in Table~\ref{tbl:IDtable} are
consistent with those obtained from head-on
collisions~\cite{Anninos98a}, mixed numerical-perturbative inspiral
mergers~\cite{2005CQGra..22S.387C} and recent fully relativistic results 
at larger separations~\cite{Baker:2006vn,Gonzalez:2006md}.  For reference, kicks from
gravitational recoil of relevance to galactic \bh{} merger
scenarios~\cite{Merritt:2004xa} have been recently estimated to second
post-Newtonian order~\cite{Blanchet:2005rj}. The kicks were found to
be dominated by the plunge phase and could reach speeds larger than
100 km/s for $0.1\le \eta \le0.24$ or $1/8\le q \le 2/3$. Recent work
using the effective one-body approach gives lower kick velocities of
at most 74\,km/s at $\eta=0.2$ or $q=0.38$~\cite{Damour:2006tr}.

\begin{figure}
\begin{center}
\includegraphics{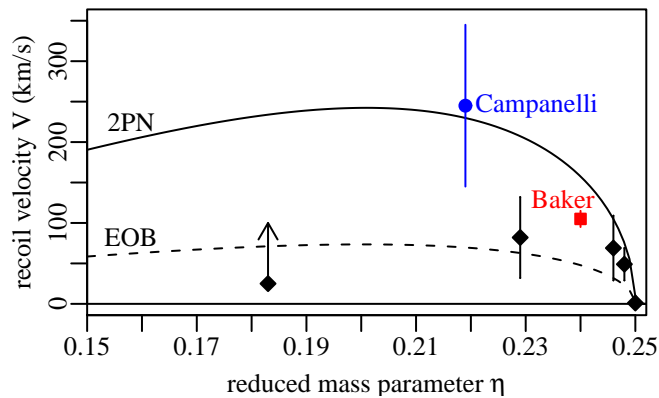}
\end{center}
\caption{A comparison of recoil velocity estimates as a function of
  $\eta$.  The 2PN estimates of Blanchet et al.~\cite{Blanchet:2005rj}
  are denoted by the solid line and those from
  Gopakumar-Damour~\cite{Damour:2006tr} by a dashed line.
  Campanelli~\cite{2005CQGra..22S.387C} and Baker et
  al.~\cite{Baker:2006vn} kicks are labeled by a circle and a box,
  respectively, and our recoil velocities are labeled with
  diamonds. Note that the smaller $\eta$ cases present lower bounds
  due to the initial data configurations studied.}
\label{fig:recoileta}
\end{figure}

In Fig.~\ref{fig:recoil_vs_time} we show the recoil velocity
accumulated as a function of time. The close detector radius manifests
itself in the non-zero initial offset of the recoil velocity of the
$q=0.32$ model. The inspiral contribution is notably absent from these
models due to their close separation (compare
Refs.\cite{Baker:2006vn,Gonzalez:2006md}). Also note that the initial
data contains an initial feature that we cannot remove as it is
already overlayed with the merger signal. In further separation models
as in Ref.~\cite{Gonzalez:2006md} the initial feature shows up in the
weaker inspiral part and can therefore easily be
removed. Fig.~\ref{fig:recoileta} shows our recoil velocities as a
function of the reduced mass parameter $\eta$, including kick
estimates found by
others~\cite{Blanchet:2005rj,Damour:2006tr,2005CQGra..22S.387C,Baker:2006vn}.
Our $q=0.85,\,\eta=0.248$ and $q=0.78,\,\eta=0.246$ cases are
comparable to the estimates from Blanchet et
al.~\cite{Blanchet:2005rj}.  The reason for the agreement is that
these cases are closer to the QC-0 equal mass case.  Thus the initial
data setup is not too far from \isco{} and quasi-circularity.  The
case $q=0.55,\,\eta=0.229$ yields the largest recoil velocity, $\sim
82$ km/s, also consistent with Blanchet et al.~\cite{Blanchet:2005rj}
who find a peak in the recoil around $q=0.4,\,\eta=0.204$.  Recently
Baker et al.~\cite{Baker:2006vn} reported a recoil velocity of
105\,km/s for $q=0.67,\,\eta=0.24$, a value larger than our kicks.
The difference is mostly due to the larger initial separation of their
\bh{s}.  Our smallest kick is obtained in the $q=0.32,\,\eta=0.183$
case.  This is likely a significant underestimation since we are
probably dealing with a \emph{single} distorted \bh{.}


\emph{Conclusions.} We have shown results from a series of unequal
mass \bbh{} inspiral simulations.  The kick velocities spanned a range
of \kickrange.  Our kick velocity estimates should be understood as a
lower bound because the initial separations are smaller than those normally
used (for example in Ref.~\cite{Blanchet:2005rj}) for the plunge
phase. The observed drop in the time to merger $t_\text{AH}$ in our
results comes from our approach in which the angular momentum of the
configuration decreases as the bare mass ratio is decreased. This
approach introduces significant uncertainty in the recoil extracted
and in particular it is possible that for our cases with $q\le0.55$, a
common event horizon already exists in the initial data slice; and,
therefore, we are evolving a single, distorted \bh{.} Ideally one
would like to compare initial data sets which are far separated and
have the same orbital frequency. We would also like to point out that
these close configurations contain significant eccentricities which
are known to influence the recoil velocities. For small eccentricities
$e\le 0.1$ Ref.~\cite{Sopuerta:2006et} shows $V_R\propto (1+e)$. More
on equal mass inspirals and eccentricities can be found in
Ref.~\cite{Buonanno:2006ui,Pfeiffer:2007yz}.

In addition to the gravitational recoil, we estimated the radiated
energy and angular momentum, paying particular attention at the energy
radiated independently in the $m=0,2$ modes.  We also monitored the
irreducible masses during the simulations, providing a good indicator
that the near-zone physics was accurately evolved. A more detailed
comparison of the dependence of the waveforms on the mass ratio from
inspirals starting outside the \isco{} is currently under
investigation.  A strong dependence of the waveform on the binary
parameters would facilitate their estimation, but at the same time it
would hinder initial detection efforts as many templates would be
required.  If waveforms, on the other hand, do not vary significantly
with mass ratios, the search effort could be easier at the expense of
accurate parameter estimation.


\emph{Acknowledgments.}  We thank E. Schnetter, M. Ansorg and J.
Thornburg for providing access to $\pi$-symmetry, initial data and
\ahz{} infrastructure, respectively, and U. Sperhake and C. Sopuerta
for helpful discussions.  Simulations performed at the CGWP, AEI, CCT,
LRZ, NCSA, NERSC, PSC, PSU and RZG.  The Center for Gravitational Wave
Physics is supported by the NSF under cooperative agreement PHY-0114375. 
Work partially supported by NCSA grant MCA02N014 and NSF
grants PHY-0244788, PHY-0354821.  PL and DS thank the hospitality of
KITP, supported by NSF PHY-9907949.



\end{document}